\documentclass[prl,twocolumn,amsmath,amssymb,superscriptaddress,showpacs] {revtex4-2}
\usepackage{hyperref}
\usepackage{graphicx,psfrag,color}
\usepackage{dcolumn}
\usepackage{bm}
\usepackage{mathbbol, appendix}
\usepackage{printlen}
\usepackage[justification=Justified]{caption}
\usepackage{diagbox}
\usepackage{amsmath}
\usepackage{pgfplots}
\usepackage{comment}
\DeclareMathOperator\erf{erf}
\usepackage[normalem]{ulem}
\usepackage{float}
\usepackage{subcaption}
\usepackage{xr-hyper}

\newcolumntype{t}[1]{D{.}{.}{#1}}

\begin{document}
\title{The Eggbox Ising Model}

\author{Mutian Shen}
\affiliation{Department of Physics, Washington University, St.
Louis, MO 63160, USA}
\affiliation{Department of Biomedical Engineering, Washington University, St.
Louis, MO 63160, USA}
\author{Yichen Xu}
\affiliation{Department of Physics, Cornell University, Ithaca, New York 14850, USA}
\author{Zohar Nussinov}
\email{corresponding author: zohar@wustl.edu}
\affiliation{Department of Physics, Washington University, St.
Louis, MO 63160, USA}
\affiliation{Rudolf Peierls Centre for Theoretical Physics, University of Oxford, Oxford OX1 3PU, United Kingdom}
\affiliation{Institut für Physik, Technische Universität Chemnitz, 09107 Chemnitz, Germany}
\begin{abstract}
    We introduce the Eggbox Ising model, a tunable construction of rugged energy landscapes defined by distances to a prescribed set of patterns. Correlated pattern ensembles realize arbitrary k-step replica-symmetry-breaking structures and controllable Parisi overlap distributions $p(q)$, consistent with the hierarchical overlap structure observed in a simple word-embedding example from empirical data. A softened variant allows a systematic expansion leading to Hopfield-type couplings (and higher-body terms). We analyze the density of states and show that suitable potentials induce discontinuous finite-temperature transitions with metastability and hysteresis.
\end{abstract}
\maketitle
\twocolumngrid

\section{Introduction}

In disordered systems such as spin glasses, the energy landscape may typically exhibit considerable complexity\cite{goldstein_viscous_2003,sciortino_potential_2005,samarakoon2016aging}. In particular, even when the ground state and the lowest energy excited state are nearly degenerate, they may be radically different in configuration space  \cite{krzakala_disorder_2005,shen_universal_2023}. Ground state searching algorithms and low-temperature Monte Carlo simulations \cite{hartmann2002optimization}  frequently encounter challenges due to the proliferation of metastable local minima. To highlight the essential intricacies of such a complex energy landscape, we introduce the ``Eggbox Ising Model.'' While the energy landscape of this model is highly complex, it  possesses a clear overall structure. The model has adjustable parameters that enable modifications to various features such as the number and depth of the local minima. This adaptability allows this model serve as a versatile arena for exploring diverse physical phenomena of numerous complex systems.

Interestingly, by fine-tuning the model parameters, we can also manipulate basic character of the Parisi overlap distribution\cite{parisi_infinite_1979,parisi_numerical_2012}. We next briefly introduce the Eggbox Ising model and explain our method for constructing Replica Symmetry Breaking (RSB) patterns of arbitrary order. Subsequently, we will demonstrate that the model harbors intriguing phase transitions. We conclude by pointing out viable connections between this model to various systems and discuss open problems.

\section{Model Description}

Most of the well-established physical models for spin glasses are constructed from 
Hamiltonians having random quenched couplings. These Hamiltonians are, generally, highly frustrated (i.e., it is impossible to find spin states that simultaneously minimize each of the many individual terms in the Hamiltonian whose minimization leads to a conflicting results). This invariably leads to an extensive number of local energy minima. This ubiquitous phenomenon  led us to introduce the bare spin glass (Eggbox Ising) model of the current work in which the
energy minima are, from the very start, directly drawn from a probability distribution. By contrast with heavily studied conventional spin glass models, the Eggbox Ising model defines the system's energy with the purpose of constructing an energy landscape of a given type. Consequently, the parameters of the Eggbox Ising model detail the different types of possible energy landscapes.

Given a set of local energy minima, the energy of a general spin configuration is a function of a Hamming type distance providing a measure of  ``how far'' this configuration is from the local energy minima.

Similar to any other Ising model, the Eggbox Ising Model is defined on the $2^{N}$  dimensional configuration space $\{-1,+1\}^N$. Here, $N$ represents the system size (the number of spins). Within this high dimensional space, $M$ configurations will be randomly chosen to be local minima. Figuratively speaking, inasmuch as energy is concerned, these local minima  configurations are like the bottoms of an eggbox carton. The value of $M$ may either be fixed (as in the current  paper), or a randomly chosen number (the latter will introduce further  disorder). The union of these $M$ local minima ($\boldsymbol{\xi}^1,\boldsymbol{\xi}^2,\dots,\boldsymbol{\xi}^{M}$)  forms the set $\mathcal{LM}$. Drawing analogies to information theory, $\mathcal{LM}$ emulates a ``Codebook''   \cite{mezard2009information}. The energies $(\epsilon_1,\epsilon_2,\dots,\epsilon_{M})$ of the $M$ local minima  are drawn from a probability distribution $p(\epsilon)$. With the energy minima characterized, we now turn to general spin configurations $\boldsymbol{\sigma}$ in $\{-1,+1\}^N$. 

In our model, the energy of such a general state is given by 
\begin{equation}\label{eq:model_energy}
E(\boldsymbol{\sigma}) = \min_\alpha [\epsilon_\alpha+V(d(\boldsymbol{\sigma},\boldsymbol{\xi}^\alpha))],
 \end{equation}
 where the minimum is over all of the $M$ local minima states $\boldsymbol{\xi}^\alpha$. The ``potential energy'' $V(d)$ is a function of a  Hamming type distance $d(\boldsymbol{\sigma},\boldsymbol{\xi}^\alpha)$ between $\boldsymbol{\sigma}$ and $\boldsymbol{\xi}^\alpha$. We set $V(0)=0$ with  $V(d)\geq 0$ and, for ease of analysis which highlights generic features of these systems, consider $V(d)=d$ and $\epsilon_\alpha = 0$ (the situation in which all of the local minima are degenerate). Such linear cost functions are germane to numerous problems in coding theory, such as ``good" low-density parity check (LDPC) codes,\cite{leverrier2015quantum,fawzi2020constant,hong2025quantum,placke2024topological}, where the energy landscape near the codeword exhibits ``linear confinement", i.e the energy of a nearby configuration of a codeword grows linearly with the Hamming distance between the configuration and the codeword.
 
 Within such a setting,  
\begin{equation}\label{eq:model_energy_simple}
E(\boldsymbol{\sigma}) =\min_\alpha d(\boldsymbol{\sigma},\boldsymbol{\xi}^\alpha).
\end{equation}
Unless otherwise specified, we will employ Eqn. (\ref{eq:model_energy_simple}) to simplify our discussion.  Ref.\cite{massen_exploring_2007} constructed a continuous (non-Ising) ``Eggbox Model'' that exhibits a similar energy landscape. Towards the end of the current work, we will turn to discuss connections between our ``Eggbox Ising model'' and this and other related models.

Complex energy landscapes with exponentially many local minima play a central role in disordered systems, from spin glasses to optimization and coding problems \cite{goldstein_viscous_2003,sciortino_potential_2005,samarakoon2016aging,hartmann2002optimization,krzakala_disorder_2005,shen_universal_2023}. Constructing simple, tunable models that allow precise control over the number, distribution, and structure of such minima can serve as both conceptual tools for understanding physical phenomena and benchmarks for testing numerical algorithms.

In this work, we introduce a novel model, which we call the \emph{Eggbox Ising Model}. This model is built to realize a rugged energy landscape with a well-controlled set of local minima. The configuration space consists of $\boldsymbol{\sigma} \in \{-1,+1\}^N$, and we explicitly specify $M$ local minima $\{\boldsymbol{\xi}^\alpha\}_{\alpha=1}^M$, which we also refer to as \emph{patterns}. The energy of a spin configuration $\boldsymbol{\sigma}$ is defined by its Hamming distance to the nearest pattern:
\begin{equation}
E(\boldsymbol{\sigma}) = \min_{\alpha} V\big(d(\boldsymbol{\sigma}, \boldsymbol{\xi}^\alpha)\big)+\epsilon^\alpha,
\label{eq:eggbox_hard_min}
\end{equation}
where $V(d)$ is a cost function satisfying $V(0) = 0$, and $d(\cdot,\cdot)$ denotes the Hamming distance, and $\epsilon^\alpha$ denotes a small perturbation associated with each local minimum.

While Eq.~\eqref{eq:eggbox_hard_min} defines a rugged landscape suitable for modeling disordered systems, its non-differentiability poses challenges for analysis and algorithm design. To overcome this, we consider a ``softened" energy landscape\cite{ramsauer2020hopfield} defined as
\begin{equation}
E_{\text{soft}}(\boldsymbol{\sigma}) = -\frac{1}{\beta_{\text{s}}} \log \sum_{\alpha=1}^M \exp\left( -\beta_{\text{s}} \big( V\big(d(\boldsymbol{\sigma}, \boldsymbol{\xi}^\alpha)\big)+\epsilon^\alpha \big) \right),
\label{eq:soft_energy_general}
\end{equation}
where $\beta_{\text{s}}$ is a softening parameter. The soft model converges to the ``hard'' model (Eq.~\eqref{eq:eggbox_hard_min}) as $\beta_{\text{s}}\to\infty$.

A particularly illuminating case arises when $V(d) = d$ and $\epsilon^\alpha = 0$, yielding
\begin{equation}
E_{\text{soft}}(\boldsymbol{\sigma}) = -\frac{1}{\beta_{\text{s}}} \log \sum_{\alpha=1}^M \exp\left( -\beta_{\text{s}} d(\boldsymbol{\sigma}, \boldsymbol{\xi}^\alpha) \right).
\end{equation}
Since $d(\boldsymbol{\sigma}, \boldsymbol{\xi}^\alpha) = \frac{N}{2} - x^\alpha$ with $x^\alpha := \frac{1}{2} \sum_i \sigma_i \xi_i^\alpha$, this is equivalent (up to constants) to
\begin{equation}
E_{\text{soft}}(\boldsymbol{\sigma}) = -\frac{1}{\beta_{\text{s}}} \log \sum_{\alpha=1}^M \exp\left( \beta_{\text{s}} x^\alpha \right).
\label{eq:soft_energy_x}
\end{equation}

Write \(Z=\sum_{\alpha=1}^{M}e^{\beta_{\mathrm{s}}x^\alpha}\) with
\(x^\alpha=\tfrac12\sum_i\sigma_i\xi_i^{\alpha}\) and use the moment
version of the Taylor (a.k.a.\ cumulant generating‑function) expansion
of \(\log Z\):
\[
\log Z
  =\log M
   +\beta_{\mathrm{s}}\langle x\rangle
   +\frac{\beta_{\mathrm{s}}^{2}}{2}\bigl(\langle x^{2}\rangle-\langle x\rangle^{2}\bigr)
   +\frac{\beta_{\mathrm{s}}^{3}}{6}\kappa_{3}
   +\mathcal{O}(\beta_{\mathrm{s}}^{4}),
\]
where \(\langle\cdots\rangle := M^{-1}\sum_{\alpha}(\cdots)\) is the
uniform pattern average and
\(\kappa_{3}=\langle (x-\langle x\rangle)^{3}\rangle\) is the third
cumulant.
Multiplying by \(-1/\beta_{\mathrm{s}}\) we obtain the softened energy
up to cubic order,
\begin{align}
E_{\mathrm{soft}}(\boldsymbol{\sigma})
  &= -\frac{\log M}{\beta_{\mathrm{s}}}
     -\langle x\rangle
     -\frac{\beta_{\mathrm{s}}}{2}\langle x^{2}\rangle
     +\frac{\beta_{\mathrm{s}}}{2}\langle x\rangle^{2}
     -\frac{\beta_{\mathrm{s}}^{2}}{6}\,\kappa_{3}
     +\mathcal{O}(\beta_{\mathrm{s}}^{3}).
\label{eq:Esoft_moment}
\end{align}

Insert \(x^\alpha\) explicitly and define the pattern
tensors\footnote{All tensors are symmetric in their indices by
construction.}
\[
h_i := \frac{1}{M}\sum_{\alpha}\xi_i^{\alpha},
\quad
J_{ij} := \frac{1}{M}\sum_{\alpha}\xi_i^{\alpha}\xi_j^{\alpha},
\quad
T_{ijk} := \frac{1}{M}\sum_{\alpha}\xi_i^{\alpha}\xi_j^{\alpha}\xi_k^{\alpha}.
\]
Then
\begin{align*}
\langle x\rangle &= \tfrac12\sum_i h_i\sigma_i,\\
\langle x^{2}\rangle &= \tfrac14\sum_{i,j}J_{ij}\sigma_i\sigma_j,\\
\kappa_{3}
  &=\langle x^{3}\rangle-3\langle x\rangle\langle x^{2}\rangle
     +2\langle x\rangle^{3}
  \\ &=\tfrac18\sum_{i,j,k}T_{ijk}\sigma_i\sigma_j\sigma_k
   - \tfrac34\Bigl(\tfrac12\sum_i h_i\sigma_i\Bigr)
               \Bigl(\tfrac14\sum_{j,k}J_{jk}\sigma_j\sigma_k\Bigr)
   \\ &+\tfrac14\Bigl(\tfrac12\sum_i h_i\sigma_i\Bigr)^{3}.
\end{align*}
Substituting these into Eq.~\eqref{eq:Esoft_moment} and grouping equal
spin monomials gives
\begin{align}
E_{\mathrm{soft}}(\boldsymbol{\sigma})
 =& -\frac{\log M}{\beta_{\mathrm{s}}}
    -\frac12\sum_i h_i\sigma_i
    -\frac{\beta_{\mathrm{s}}}{8}\sum_{i,j}J_{ij}\sigma_i\sigma_j
\nonumber\\
 &\quad
    +\frac{\beta_{\mathrm{s}}}{8}\Bigl(\sum_i h_i\sigma_i\Bigr)^{2}
    -\frac{\beta_{\mathrm{s}}^{2}}{48}\sum_{i,j,k}T_{ijk}\sigma_i\sigma_j\sigma_k
    +\mathcal{O}(\beta_{\mathrm{s}}^{3}).
\label{eq:Esoft_full}
\end{align}
The \(\bigl(\sum_i h_i\sigma_i\bigr)^{2}\) piece can be
absorbed into a renormalised quadratic coupling
\(J_{ij}\!\to\!J_{ij}-h_i h_j\) if desired, but is written out
explicitly here to emphasise the distinct origins of the two quadratic
contributions.

If every pattern \(\boldsymbol{\xi}_{\alpha}\) is stored together with
its negative \(-\boldsymbol{\xi}_{\alpha}\) ("pair
symmetry''), then \(h_i=0\) for all \(i\).
Eq.~\eqref{eq:Esoft_full} reduces to
\begin{eqnarray}
E_{\mathrm{soft}}(\boldsymbol{\sigma})
  =&& -\frac{\log M}{\beta_{\mathrm{s}}}
    -\frac{\beta_{\mathrm{s}}}{8}\sum_{i,j}J_{ij}\sigma_i\sigma_j \nonumber
    \\
   && -\frac{\beta_{\mathrm{s}}^{2}}{48}\sum_{i,j,k}T_{ijk}\sigma_i\sigma_j\sigma_k
   +\mathcal{O}(\beta_{\mathrm{s}}^{3}).
\label{eq:Esoft_pairsym}
\end{eqnarray}
When small perturbations $\epsilon^\alpha$ are included in the softened model via
\(
\omega^\alpha := x^\alpha - \epsilon^\alpha,
\)
the cumulant expansion gains additional contributions beyond those from $x^\alpha$ alone. The energy in Eq.~\eqref{eq:Esoft_pairsym} receives the following corrections:
\begin{align}
\langle \omega \rangle - \langle x \rangle &= -\langle \epsilon \rangle, \\
\langle \omega^2 \rangle - \langle x^2 \rangle &=   - 2 \langle \epsilon x \rangle+\langle \epsilon^2 \rangle, \\
\langle \omega^3 \rangle - \langle x^3 \rangle &=  - 3 \langle \epsilon x^2  \rangle + 3 \langle \epsilon^2 x \rangle - \langle \epsilon^3 \rangle.
\end{align}
For simplicity, we set $\langle \epsilon \rangle = 0$. Defining
\[
\delta h_i := \frac{1}{M} \sum_\alpha \epsilon^\alpha \xi_i^\alpha,
\quad
\delta J_{ij}^{(\epsilon)} := \frac{1}{M} \sum_\alpha 
 \epsilon^\alpha\, \xi_i^\alpha \xi_j^\alpha ,
\]
then up to the first order of $\epsilon^\alpha$, the full energy 
becomes
\begin{eqnarray}
E_{\mathrm{soft}}(\boldsymbol{\sigma})
  =   &&-\frac{\log M}{\beta_{\mathrm{s}}}
     -\frac{\beta_{\mathrm{s}}}{2} \sum_i \delta h_i \sigma_i \nonumber
     \\
    && -\frac{\beta_{\mathrm{s}}}{8}\sum_{i,j}(J_{ij}-\beta_{\mathrm{s}}\delta J_{ij}^{(\epsilon)})\sigma_i\sigma_j \nonumber
    \\ &&    -\frac{\beta_{\mathrm{s}}^{2}}{48}\sum_{i,j,k}
         T_{ijk}  
        \sigma_i\sigma_j\sigma_k
     + \mathcal{O}(\beta_{\mathrm{s}}^3).
\label{eq:Esoft_pairsym_eps}
\end{eqnarray}

This result shows that a small perturbation $\epsilon^\alpha$ to the valley depth of each pattern manifests, at leading order, as an effective external field $\delta h_i$ acting on each spin and a correction $\delta J_{ij}^{(\epsilon)}$ to the Hopfield-type two-body couplings.

\section{From $1$-RSB to $k$-RSB Structure}

In what follows, we provide a succinct review  of the concept of Replica Symmetry Breaking (RSB). In one-step replica symmetry breaking, or $1$-RSB, the entire configuration space is segmented into multiple subspaces (referred to as `pure states'). Within these subspaces, replicas are symmetric under any permutation, meaning that for replicas $\alpha$ and $\beta$ within the same pure state, their overlap $q_{\alpha\beta}$ equals $q_1$, irrespective of their specific values. Conversely, pure states themselves are symmetric under permutation, such that for replicas $\alpha$ and $\beta$ belonging to different pure states, their overlap $q_{\alpha\beta}$ equals $q_0$, regardless of their values. For higher orders of RSB, the overlap between replicas can assume a broader array of values. These values adhere to a fractal structure\cite{568530b2-62d5-3d43-9e2f-adfdf424006b} akin to the one depicted in the Fig. \ref{fig:fractalrsb}(a). Furthermore, the structure of Replica Symmetry Breaking (RSB) exhibits ultrametricity\cite{parisi_origin_2000}. When randomly selecting three replicas, denoted as $\alpha$, $\beta$, and $\gamma$, the overlaps among them, represented as $q_{\alpha\beta}, q_{\alpha\gamma}, q_{\beta\gamma}$, can result in only two possible scenarios. These are either $q_{\alpha\beta} = q_{\alpha\gamma} = q_{\beta\gamma}$, indicating equal overlap among all three replicas, or $q_{\alpha\beta} > q_{\alpha\gamma} = q_{\beta\gamma}$, signifying that one pair of replicas has a greater overlap than the other two pairs. Note in the ultrametric space, $q_{\alpha\beta} > q_{\alpha\gamma} > q_{\beta\gamma}$ is impossible.

Next, we will discuss the RSB structure in the Eggbox Ising model.

\subsection{$1$-RSB}

In the Eggbox Ising model, the local minima $\boldsymbol{\xi}^1,\boldsymbol{\xi}^2,\dots, \boldsymbol{\xi}^M$ naturally induce the corresponding pure states with replica  indices $\alpha=1,2,\dots, M$. With $\mathcal{LM}$ a set of randomly {\it uncorrelated} selected spin configurations in the configuration space $\{-1,+1\}^N$, the zero-temperature Parisi overlap 
\begin{equation}
\begin{split}
    q_{\alpha\beta}&\equiv \frac{1}{N}\sum_{i=1}^N\langle s_i\rangle_\alpha\langle s_i \rangle_\beta
    \stackrel{T=0}{=} 
    \frac{1}{N}\sum_{i=1}^N\xi_{\alpha,i} \xi_{\beta,i}  \\
    &\stackrel{N\to \infty}{=}  \begin{cases}
        1, \quad &\alpha=\beta,\\
        0, \quad &\alpha\neq\beta
    \end{cases}
\end{split}
\end{equation}
Here, $\langle \cdot \rangle_\alpha$ denotes the thermal average for replica $\alpha$, and $\xi_{\alpha,i}$ denotes the $i$-th spin of $\boldsymbol{\xi}^\alpha$. Note that this 
situation of uncorrelated minima aligns with one-step replica symmetry breaking. Consequently, we refer to Eqn. (\ref{eq:model_energy},\ref{eq:model_energy_simple}) as the 1-RSB Eggbox Ising model, when $\mathcal{LM}$ constitutes a collection of $M$ independently randomly selected spin configurations within the space $\{-1,+1\}^N$.

\subsection{$k$-RSB}

The above construct can be readily extended to an arbitrary step Replica Symmetry Breaking (RSB). To illustrate, the construction of a 2-RSB model from a 1-RSB model involves such a systematic procedure:

\begin{enumerate}
\item Select a specific $\boldsymbol{\xi}^\alpha$ from the set $\mathcal{LM}$.
\item Fix a randomly chosen half of the spins within $\boldsymbol{\xi}^\alpha$, and resample the remaining spins for $c$ times, resulting in $c$ new configurations: $\boldsymbol{\xi}^{\alpha 1}, \boldsymbol{\xi}^{\alpha 2}\dots \boldsymbol{\xi}^{\alpha c}$.
\item Remove the original $\boldsymbol{\xi}^\alpha$ from $\mathcal{LM}$, replacing it with $\boldsymbol{\xi}^{\alpha 1}, \boldsymbol{\xi}^{\alpha 2}\dots \boldsymbol{\xi}^{\alpha c}$. This step increases the number of local minima by a factor of $c$.
\item Apply steps 1 to 3 iteratively for every index $\alpha$ within the model.
\end{enumerate}

\begin{figure}[!htb]
    \centering
    \includegraphics[width =0.42\textwidth]{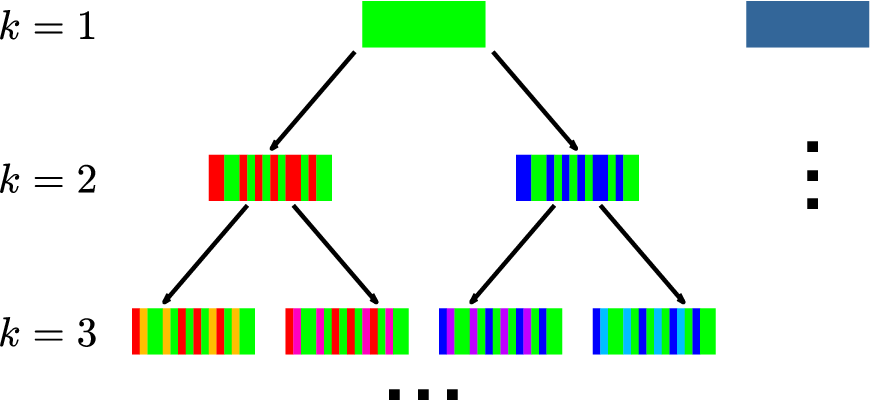}
    \includegraphics[width =0.4\textwidth]{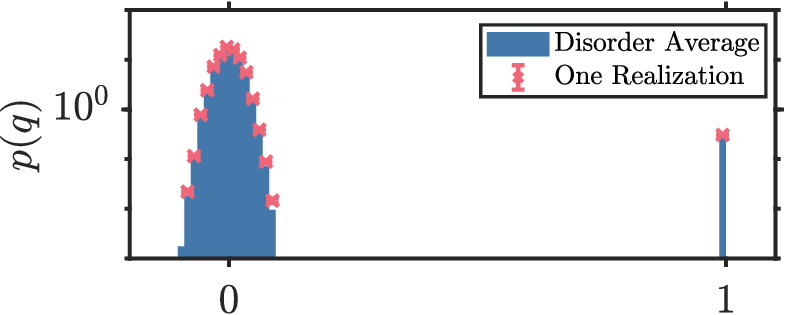}
    \includegraphics[width=0.4\textwidth]{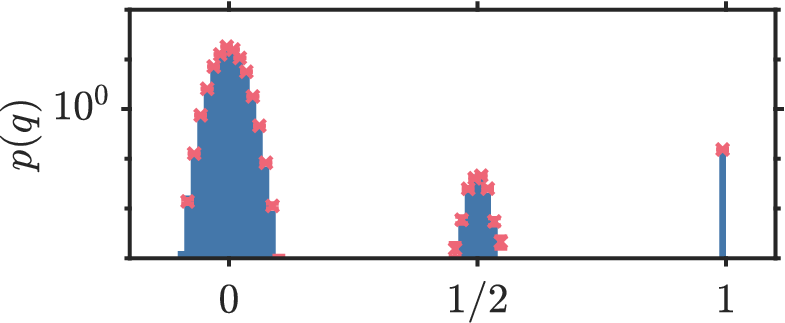}
    \includegraphics[width=0.4\textwidth]{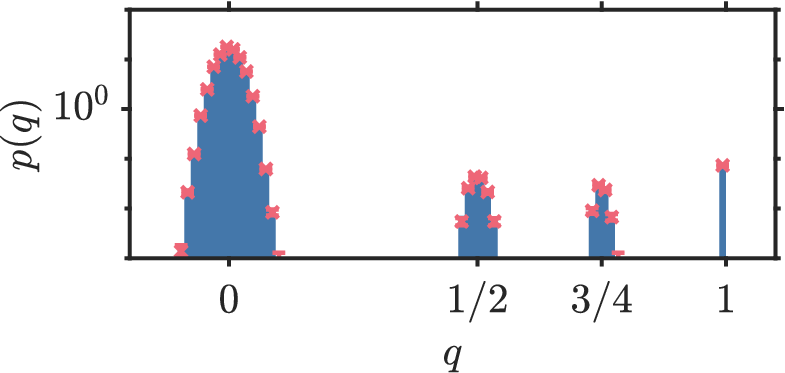}
    \caption{Iterative evolution from the $1$-RSB to the $k$-RSB Eggbox Ising model. (a) Initially ($k=1$), each $\boldsymbol{\xi}^\alpha$ undergoes random spin sampling, represented in green for clarity in subsequent stages; this process is equally applicable to other $\boldsymbol{\xi}^\beta$'s (navy blue configuration on the right). In advancing to the next ($k=2$) level, half of the spins remain unchanged with the remainder being resampled (red and blue spins). Progressing further to the third level ($k=3$), this bifurcation is repeated, with half of either the red or blue spins being held fixed while the others are  resampled. (b) The Parisi overlap distribution $P(q)$ for $k=1,2,3$ (left to right) simulated on $100$ disorder realizations with $N=2048, M_0=256$ using $500,000$ samples for each disorder realization. The red traces show the Parisi overlap distribution for one realization; this histogram is almost identically covered the disorder-averaged Parisi overlap distribution. The equivalence between the latter two  distributions illustrates the self-averaging for disorder realizations that are generated via the same process with fixed parameters ($M_0,c,k$).}
    \label{fig:fractalrsb}
\end{figure}

This methodical approach facilitates the generation of higher-order RSB within the Eggbox Ising model. An illustration of this process is provided in Fig. \ref{fig:fractalrsb}(a). 

In the progression to higher step Replica Symmetry Breaking (RSB), specifically transitioning from a $k$-RSB to a $(k+1)$-RSB Eggbox Ising model, it is essential to maintain all previously fixed spins. Subsequently, one should proceed to resample a randomly chosen half of the remaining unfixed spins. This is followed by resampling for $c$ times, wherein the original configuration is replaced by $c$ newly generated spin configurations. In what follows, we set $c=2$ in the described numerical simulations.

For $k$-RSB Eggbox Ising model, there are $M=M_0c^{k-1}$ local minima, where $M_0$ denotes the size of $\mathcal{LM}$ before any `split'. Also, $q_{\alpha\beta}\in\{0,\frac{1}{2},\frac{3}{4},\frac{7}{8},\dots,1-(\frac{1}{2})^{k-1},1\}$, see Fig. \ref{fig:fractalrsb}(b). If every disorder realization follows the same generation procedure, then the self-averaging property of the overlap distribution becomes evident. This is also demonstrated in Fig. \ref{fig:fractalrsb}(b). As a result, for subsequent calculations of the overlap distribution, it may suffice to use just a single disorder realization. However, we caution that if randomness is inherent in the parameters $M_0, c$, and $k$, the self-averaging property of the overlap distribution is no longer apparent. This issue might itself be an interesting subject for further research.

The local minima generated through such a procedure evidently display the fractal natural of RSB. The ultrametricity of RSB is also straightforward to demonstrate. We denote $k_{\alpha\beta}$ as the layer number of the `lowest common ancestor' for replicas $\alpha$ and $\beta$, where the `lowest common ancestor' is defined as the first common ancestor when tracing the lineage upwards from $\alpha$ and $\beta$. This $k_{\alpha\beta}$ determines the overlap $q_{\alpha\beta}$. If $k_{\alpha\beta}$ equals $k_{\alpha\gamma}$, indicating that $\alpha$, $\beta$, and $\gamma$ share the same lowest common ancestor, then it follows that $q_{\alpha\beta}=q_{\alpha\gamma}=q_{\beta\gamma}$. Conversely, if $k_{\alpha\beta}>k_{\alpha\gamma}$, this suggests that the lowest common ancestor of $\alpha$ and $\gamma$ is situated above that of $\alpha$ and $\beta$, naturally resulting in $q_{\alpha\beta}> q_{\alpha\gamma} = q_{\beta\gamma}$ .

Here we point out that the resampling in the splitting procedure from $k$-RSB to $(k+1)$-RSB is highly tunable. For instance, to increase the probability of obtaining $q_{\alpha\beta}=\frac{1}{2}$, one might adopt an `unbalanced' split approach. This involves dividing a specific $\boldsymbol{\xi}^\alpha$ into a greater number of variants $c_\alpha$ compared to other $\boldsymbol{\xi}^\beta$'s, namely $c_\alpha$ is not a constant $c$ anymore and $c_\alpha>c_\beta$.  Consequently, this increases the likelihood of sampling $\alpha_{i_1},\alpha_{i_2}$ pairs sharing the same parent. Also, the ratio of spins resampled can be tuned to induce any $q_{\alpha\beta}$ values, other than $1-(\frac{1}{2})^{k-1}$. In essence, this implies that within the Eggbox Ising model, the Parisi overlap distribution can be modulated. 

By iterative (generally non-uniform) resampling of the general levels, the reader can readily see how {\it our construct enables the construction of general overlap distributions $P(q)$ that are not necessarily of the simplest symmetric $k-$RSB type}.

\begin{figure}[!htb]
    \centering
    \includegraphics[width=0.48\textwidth]{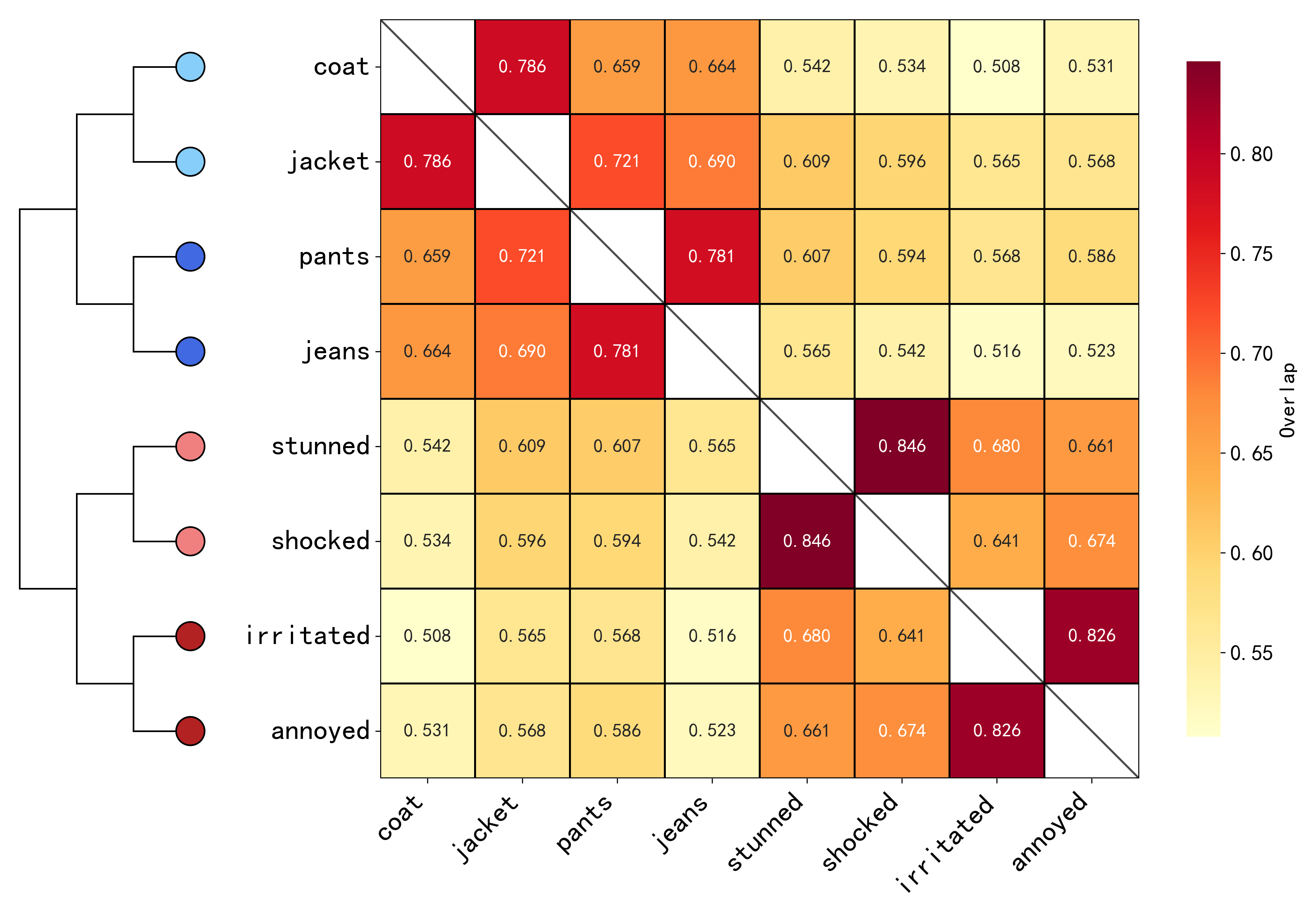}
    \caption{An illustrative overlap matrix computed from binarized word embeddings for eight English words. The matrix exhibits a hierarchical structure: two coarse semantic clusters (clothing vs.\ emotions) and finer sub-clusters within each group. This mirrors the multi-level organization of overlaps encountered in $k$-RSB structures.}
    \label{fig:nlp_overlap}
\end{figure}

Interestingly, a closely related \emph{hierarchical} overlap structure can be observed in empirical data beyond spin glasses. As a simple illustration, consider a set of eight English words (e.g., \texttt{coat, jacket, pants, jeans, stunned, shocked, irritated, annoyed}) and represent each word by a (pretrained) continuous embedding vector $\mathbf{e}_a \in \mathbb{R}^D$. In our implementation we use the HuggingFace \texttt{Transformers} library with the pretrained BERT model \texttt{bert-base-uncased}\cite{devlin2019bert,GooglebertBertbaseuncasedHugging2024}, for which the hidden dimension is $D=768$. For each word, we apply the model's WordPiece tokenizer, take the last-layer hidden states, and use mean pooling over tokens to obtain $\mathbf{e}_a$. We then binarize by taking the element-wise sign,
\begin{equation}
\mathbf{s}_a := \mathrm{sign}(\mathbf{e}_a), \qquad \mathbf{s}_a \in \{-1,+1\}^D,
\end{equation}
Empirically, such aggressive quantization/binarization is often found to have only a minor impact on downstream performance while providing a convenient discrete representation \cite{shakir2024quantization}.
and define an ``overlap'' between words by
\begin{equation}
q_{ab} := \frac{1}{D}\, \mathbf{s}_a \cdot \mathbf{s}_b.
\end{equation}
then the resulting overlap matrix (Fig.~\ref{fig:nlp_overlap}) displays a clear hierarchical organization: at the coarsest level the words split into two semantic groups (clothing vs.\ emotions), and within each group there is a further refinement (e.g., outerwear vs.\ trousers within clothing). This provides a concrete, intuitive motivation for constructing patterns that \emph{themselves} carry an explicit multi-level (ultrametric) overlap structure in our Eggbox Ising model.

\section{Density of States}\label{sec:dos}

To set the ground for our analysis of the phase transitions, we discuss the density of states that is associated with the energy landscape. The partition function of any system,  
\begin{equation}
    Z=\sum_E \Omega(E)e^{-\beta E}=2^N\int^{+\infty}_{-\infty} \omega(E)e^{-\beta E}dE.
\end{equation}
Here, $\Omega(E)$ is the number of states/configurations of energy $E$, and $\omega(E)$ is the fraction of states having energy $E$ with $\int^{+\infty}_{-\infty} \omega(E)dE=1$. 

In the 1-RSB Eggbox Ising model, the energy $E(\boldsymbol{\sigma})$ of a configuration $\boldsymbol{\sigma} \in \{-1,+1\}^N$ is fundamentally determined by the minimum distance from $\boldsymbol{\sigma}$ to $M$ randomly selected spin configurations $\{\boldsymbol{\xi}^\alpha\}$, according to Eq. (\ref{eq:eggbox_hard_min}). It is apparent that the normalized density of states, denoted as $\omega(E)$, can be inferred from the distribution of local minima by calculating the extreme value distribution\cite{gut2006probability}. To write down its form explicitly, we denote the standard normal distribution function as $g(x) = (\frac{1}{\sqrt{2\pi}})\exp(-\frac{x^2}{2})$, and its cumulative distribution function as $G(x) = \int^x_{-\infty} g(t)dt = \frac{1}{2}[1+\erf(\frac{x}{\sqrt{2}})]$, where $\erf(x)$ represents the error function. Importantly, the distance between two configurations randomly sampled in $\{-1,+1\}^N$ follows the distribution $g(\frac{x-\frac{N}{2}}{\frac{\sqrt{N}}{2}})$, by the central limit theorem \cite{gut2006probability}. Hence, using the energy expression in Eq. (\ref{eq:eggbox_hard_min}), $\omega(E)$ is given by:

\begin{align}
\label{eq:1rsbdos}
    \omega(E) =& \frac{M}{\sigma}\left[1-G\left(\frac{E-\mu}{\sigma}\right)\right]^{M-1}g\left(\frac{E-\mu}{\sigma}\right).\nonumber \\
    =&\frac{M}{\sigma}\left[1-G(E')\right]^{M-1}g(E').
\end{align}
Where $E'=\frac{E-\mu}{\sigma}, \mu=N/2,\sigma=\sqrt{N}/2$. Intuitively, the term $\left[1-G(E')\right]^{M-1}$ is derived from the extreme value distribution, guaranteeing $E'$ as the minimal energy to some given local minima compared to the energies computed with respect to other local minima. According to Fisher-Tippett-Gnedenko theorem\cite{embrechts2013modelling}, the approximate center of this extreme value distribution deviates from the original mean in the form of $\sim\sqrt{\ln M}$.

For cases where $k \geq 2$, the expression of $\omega(E)$ becomes increasingly intricate due to the correlations present among the $\boldsymbol{\xi}^\alpha$ configurations. More explicitly, the difficulty lies in iteratively tracking the distribution of relative distance to newly generated energy minima for $k + 1$ from the minima of $k$. This complexity can be reduced to evaluating the distribution of a variable $U$, defined as $U=\min\{X+Y,X+Z\}$, where $X, Y, Z$ are independent random variables. In reality, $X$ represents the sub-configuration of the spins that are going to be fixed, and $Y,Z$ correspond to the newly sampled sub-configurations. For simplicity, let us assume that $X, Y, Z$ are all 
governed by a probability 
density 
$g(x)$ (with cumulative distribution $G(x) \equiv \int^x_{-\infty} g(t)dt$). The cumulative distribution function of $U$ is thus 
\begin{equation}
    F_U(x)=P(U\leq x) = 1-\int^{+\infty}_{-\infty}g(y)(1-G(x-y))^2 dy.
\end{equation}

\begin{figure}[!htb]
    \centering
    \includegraphics[width=0.46\textwidth]{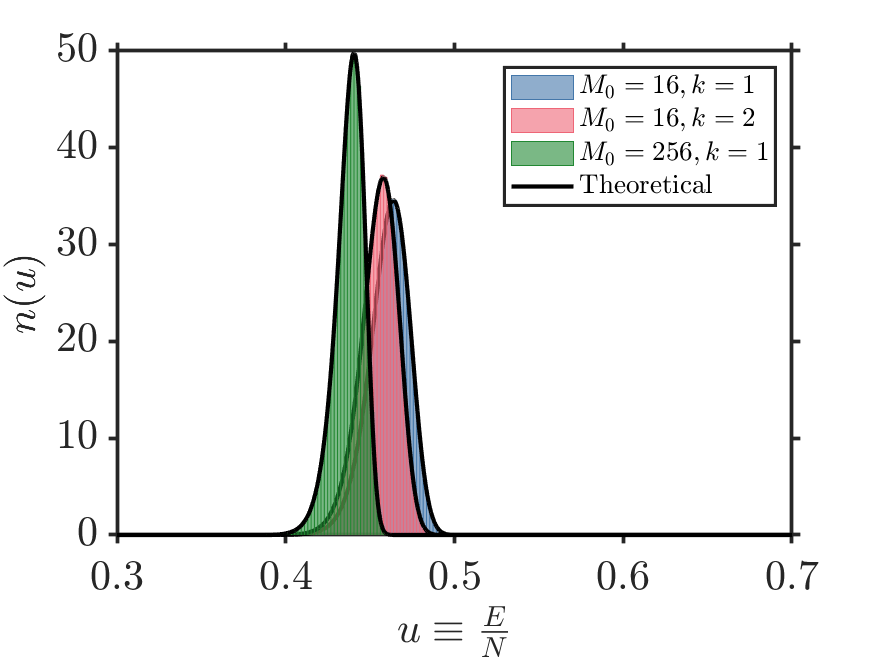}
    \caption{Normalized density of states $n(u)\equiv N\omega(Nu)$ versus the energy density $u\equiv \frac{E}{N}$ with $\int_0^\infty n(u)du= \int_0^\infty \omega(Nu)d(Nu)\overset{E=Nu}{=}1$. For each $M_0$ and each $k$, the $500,000$ configurations were sampled from a $N=512$ system. From right to left, we have $M_0=16,k=1$; $M_0=16,k=20$; $M_0=256,k=1$. The histograms for varying $M_0$ and $k$ match the theoretical curves.}
    \label{fig:dos}
\end{figure}

We examine these by comparing our theoretical computation using the procedure above and numerical simulation in Fig. \ref{fig:dos}.

It is helpful to intuitively summarize the characteristics of the density of states in the Eggbox Ising model. When $M$ is finite and $N \to \infty$, the width of $g(E')$ tends towards zero. This convergence results in the density of states becoming concentrated around $u\equiv E/N = \frac{1}{2}$. This particular scenario facilitates subsequent analytical endeavors. However, it is crucial to remain cognizant of the fact that when $M$ is sufficiently large, the density of states will disperse towards $u < \frac{1}{2}$ (intuitively, as the number of local minima increases, the shortest distance between a randomly chosen configuration and these local minima is naturally more likely to be smaller), a phenomenon attributable to the characteristics of the extremal value distribution.

\section{Phase Transitions}

\subsection{Maximal Probability at Equilibrium}

Consider the free energy $F(d)$ of the configurations with distance $d<N/2$ to a local minima $\boldsymbol{\xi}^\alpha$ sufficiently small that the thermal fluctuation from $\boldsymbol{\xi}^\alpha$ to any other minima will be strongly suppressed. In this case, we have:
\begin{equation}
\label{Fd}
    F(d) = E(d)-TS(d) = E(d)-\frac{1}{\beta}\ln \Omega(d)
\end{equation}
Here $\Omega(d) = 2^{N} \omega(E=d)$
(with $\omega(E)$ of Eqn. (\ref{eq:1rsbdos})) is the number of possible spin configurations. For small $d$ such that only a single local minimum can be considered in Eqn. (\ref{eq:1rsbdos}) (i.e., when the $(M-1)$ factors of $(1-G(E'))$ in the second term of Eqn. (\ref{eq:1rsbdos}) may be set to unity), $\Omega(d)=\frac{N!}{d!(N-d)!}$. In what follows, $d_*\equiv \arg\min\limits_{d}F(d)$.  Since it is an extremum, the free energy minimum must trivially satisfy 
\begin{align}
\label{FE}
    0 \simeq&  F(d^*+1)-F(d^*) \nonumber\\
    =&  E(d^*+1)-E(d^*) -\frac{1}{\beta}\ln\Big(\frac{N-d^*}{d^*+1}\Big).
\end{align}
When $E(d)=V(d) = d$,
\begin{equation}\label{eq:freeminima}
    d_* \simeq \frac{N}{1+e^{\beta}}.
\end{equation}

As a most basic check, for infinitely large temperature $\beta = 0$, meaning there is no energy cost for flipping, it follows that $d_*\simeq \frac{N}{2}$. Hence, for the simple case of $V(d) = d$ and sufficiently small $M$, most of the spin configurations are condensed around $d_*=\frac{N}{2}$, which aligns with our intuitive expectations.

In fact, this method can be very naturally and easily extended to more well-known models, such as the Curie-Weiss model. The Curie-Weiss model describes the energy of an all-to-all coupled Ising system as:
\begin{equation}
E = -\frac{J}{2N} \sum_{i,j} \sigma_i \sigma_j,
\end{equation}
where $\sigma_i \in \{ \pm 1 \}$ and $J$ is the interaction strength. Using the identity $\sum_{i,j} \sigma_i \sigma_j = \left( \sum_i \sigma_i \right)^2$, the energy can be rewritten as:
\begin{equation}
E = -\frac{J}{2N} \left( \sum_i \sigma_i \right)^2 + \frac{J}{2}.
\end{equation}

Let $d$ be the number of spins with $\sigma_i = +1$, then $\sum_i \sigma_i = N - 2d$, and the energy becomes:
\begin{equation}
E(d) = -\frac{J}{2N} (N - 2d)^2 + \frac{J}{2}.
\end{equation}

Following the same approach as in Eqn. (\ref{FE}), we computed the energy difference:
\begin{align}
    \nonumber E(d^*+1)-E(d^*) &= -\frac{J}{2N}[(N-2(d^*+1))^2 - (N-2d^*)^2] \\ 
&= \frac{2J}{N}(N-2d^*).
\end{align}

Setting the free energy difference to zero and solving for $d^*$:
\begin{equation}
\frac{2J}{N}(N-2d^*) = \frac{1}{\beta}\ln\Big(\frac{N-d^*}{d^*+1}\Big).
\end{equation}

At the critical point where $d^* = N/2$, this equation becomes:
\begin{equation}
0 = \frac{1}{\beta_c}\ln(1) = 0,
\end{equation}
which is satisfied for any finite $\beta_c$. However, the critical behavior emerges when we consider the \emph{stability} of this stationary point. Since $d$ is discrete, we characterize stability by the discrete second difference (the lattice analogue of the second derivative)
\begin{equation}
\Delta^2 F(d) \equiv F(d+1)-2F(d)+F(d-1).
\end{equation}
A stationary point $d_*$ satisfying $F(d_*+1)-F(d_*)\simeq 0$ corresponds to a stable local minimum if $\Delta^2F(d_*)>0$, and to an unstable extremum if $\Delta^2F(d_*)<0$.

For
\begin{equation}
F(d)=E(d)-\frac{1}{\beta}\ln\binom{N}{d},
\end{equation}
the discrete curvature separates into energetic and entropic contributions,
\begin{equation}
\Delta^2F(d)=\Delta^2E(d)-\frac{1}{\beta}\Delta^2\ln\binom{N}{d},
\end{equation}
with $\Delta^2E(d)=E(d+1)-2E(d)+E(d-1)$.
Using the binomial identities
\[
\frac{\binom{N}{d+1}}{\binom{N}{d}}=\frac{N-d}{d+1},
\qquad
\frac{\binom{N}{d}}{\binom{N}{d-1}}=\frac{N-d+1}{d},
\]
we obtain
\begin{align}
\Delta^2\ln\binom{N}{d}
&=\left[\ln\binom{N}{d+1}-\ln\binom{N}{d}\right] \nonumber\\
&\qquad -\left[\ln\binom{N}{d}-\ln\binom{N}{d-1}\right] \nonumber\\
&=\ln\!\left(\frac{N-d}{d+1}\right)-\ln\!\left(\frac{N-d+1}{d}\right) \nonumber\\
&=\ln\!\left(\frac{d(N-d)}{(d+1)(N-d+1)}\right).
\end{align}
Therefore,
\begin{equation}
\Delta^2F(d)
=\Delta^2E(d)
-\frac{1}{\beta}
\ln\!\left(\frac{(d+1)(N-d+1)}{d(N-d)}\right).
\label{eq:disc_curvature_general}
\end{equation}
For the Curie--Weiss energy in Eq.~(380), which is quadratic in $d$, the energetic contribution to the discrete curvature is a constant,
\begin{equation}
\Delta^2E(d)=-\frac{4J}{N}.
\end{equation}
At the symmetric point $d=N/2$ (for even $N$), the entropic term is positive and competes with the negative energetic curvature. To make the large-$N$ expansion explicit, we evaluate the logarithm in Eq.~(\ref{eq:disc_curvature_general}) at $d=N/2$,
\begin{align}
\nonumber \ln\!\left(\frac{(d+1)(N-d+1)}{d(N-d)}\right)\Bigg|_{d=N/2}
&=\ln\!\left(\frac{(\frac{N}{2}+1)^2}{(\frac{N}{2})^2}\right)\\
&=2\ln\!\left(1+\frac{2}{N}\right),
\end{align}
Using $\ln(1+x)\simeq x$ with $x=2/N$, and substituting this together with $\Delta^2E(d)=-4J/N$ into Eq.~(\ref{eq:disc_curvature_general}) yields
\begin{equation}
\Delta^2F\!\left(\frac{N}{2}\right)
\simeq \frac{4}{N}\left(\frac{1}{\beta}-J\right),
\end{equation}
showing that the curvature changes sign at the critical inverse temperature
\begin{equation}
\beta_c = \frac{1}{J}.
\end{equation}
This sign change signals the loss of stability of the symmetric free-energy minimum and the onset of spontaneous symmetry breaking.

For a nearest-neighbor Ising model with coordination number $z$, the mean-field energy is approximated by:
\begin{equation}
E_{\text{mean}} = -\frac{Jz}{2N} (N - 2d)^2.
\end{equation}
This resembles the Curie-Weiss form with an effective coupling strength $J_{\text{eff}} = Jz$, leading to a critical inverse temperature:
\begin{equation}
\beta_c = \frac{1}{zJ}.
\end{equation}


It is worth noting that the free energy analysis method employed here to determine phase transition temperatures is mathematically equivalent to the traditional approach of expressing the free energy as a function of order parameters (such as magnetization) and then performing minimization analysis. However, we believe that the analysis through the Eggbox model provides a clearer physical picture that, at certain levels, can be more intuitive than traditional analytical approaches. 

\subsection{Phase Transitions for Various Potentials}

The Eggbox Ising model lucidly exemplifies the connection between the energy landscapes and their associated spin-glass phase transitions. In what follows, we will viable phase transitions that appear for different functions $V(d)$. We have studied these phase transitions by both  numerical calculations and analytical considerations. The numerical computations (including our reported results for the internal energy density $u$ and the Parisi Overlap $P(q)$) employed the standard Monte Carlo method based on the Metropolis Hasting algorithm\cite{robert_metropolis-hastings_2016}.

\begin{figure}
    \centering
    \includegraphics[width=0.40\textwidth]{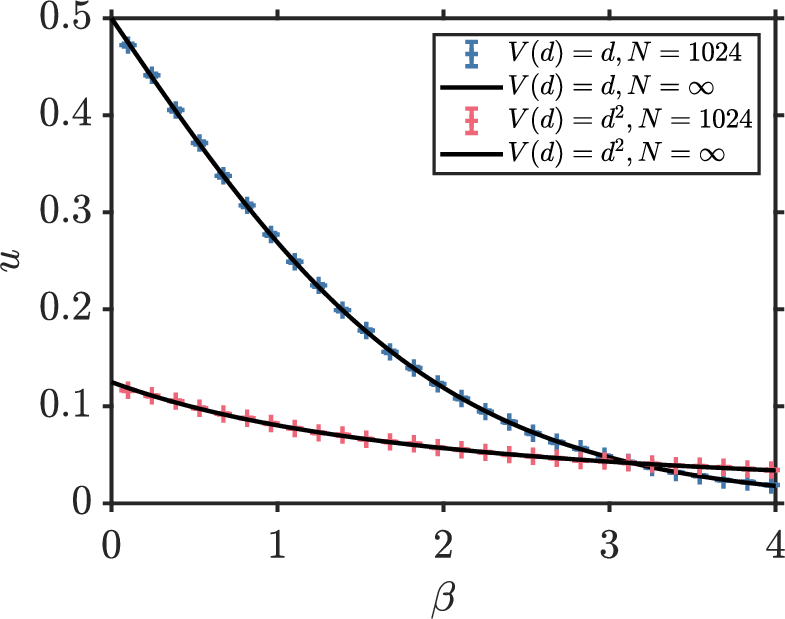}
    \caption{There is no finite temperature phase transition for $V(d)=d$ and $V(d)=d^2$. Here,  we plot the internal energy density (energy per spin) as a function of the inverse temperature $\beta$. The displayed data were obtained from configurations that were sampled across systems of size $N=1024$, while fixing $M=2$ and the ratio $a\equiv \frac{d_0}{N}=\frac{1}{4}$. To ensure thermalization, at every temperature, each sample of a single disorder realization underwent $110$ Metropolis-Hasting sweeps from random configurations. Each sweep includes $N$ flipping attempts. Subsequently, the internal energy density was calculated for $100$ samples. For the thermodynamic limit $N=\infty$, the internal energy was determined through theoretical computations (Eqns. (\ref{eq:freeminima}, \ref{eq:piecewise})).}
    \label{fig:no_phaset}
\end{figure}

\begin{figure}[!htb]
    \centering
    \includegraphics[width=0.45\textwidth]{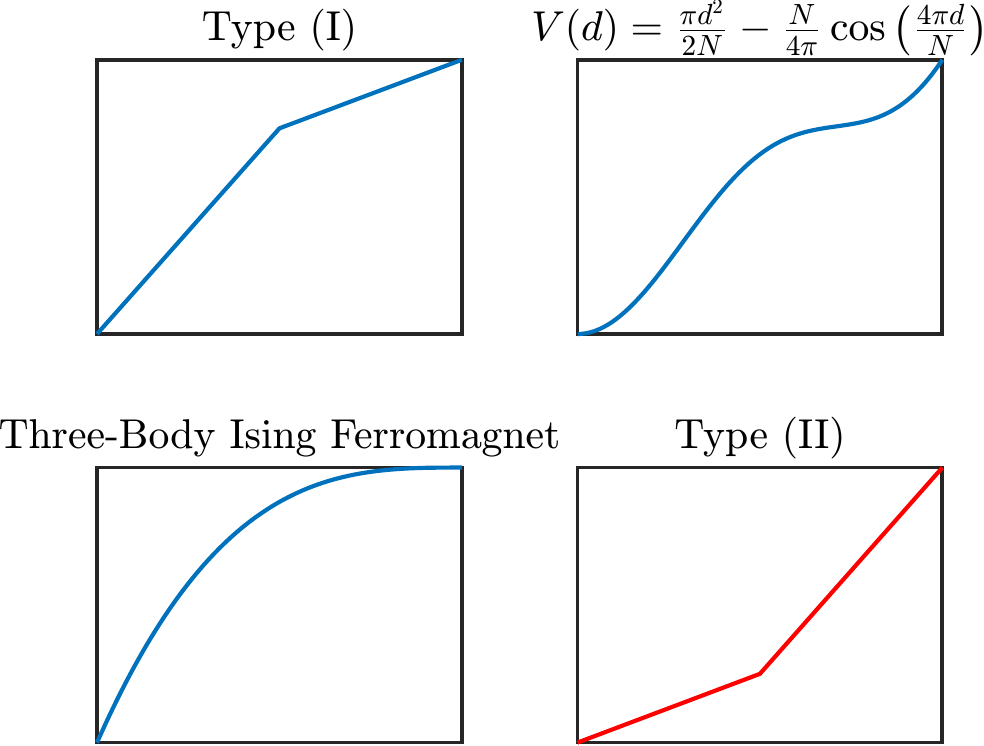}
    \caption{Sketches for the examples of potential $V(d)$. From left to right: the first and second ones are both piecewise linear potentials (Eqn. (\ref{eq:piecewise})) with $\gamma<1$ (type (I)) and $\gamma>1$ (type (II)) respectively. The third one $V(d)= \frac{\pi d^2}{2N}-\frac{N}{4\pi}\cos\left( \frac{4\pi d}{N}\right)$ is more analytical, behaving similarly as type (I) piecewise linear potential in terms of the phase transition.}
    \label{fig:potentials}
\end{figure}

One pathway for engineering finite temperature phase transitions is that of generating systems displaying multiple free energy minima. Simple potentials such as like $V(d)=d$ or $V(d)=d^2$ do not give rise to finite temperature phase transitions due to the absence of multiple free energy minima, see Fig. \ref{fig:no_phaset}. However, by introducing a modicum of complexity, systems governed by more nontrivial $V$ (for example, $V(d)= \frac{\pi d^2}{2N}-\frac{N}{4\pi}\cos\left( \frac{4\pi d}{N}\right)$ (the rightmost panel of  Fig. \ref{fig:potentials})) do exhibit finite temperature phase transitions. For ease of analysis, we start off by considering a piecewise linear potential, 
\begin{equation}\label{eq:piecewise}
    V(d)=\begin{cases}
        d, &d\leq d_0,\\
        d_0+\gamma (d-d_0),& d > d_0,
    \end{cases}
    \\
\end{equation}
with $d_0<\frac{N}{2}$. There are two qualitatively different types of such piecewise linear potentials- those with $\gamma<1$ and those for which $\gamma>1$ (the leftmost and central panel of Fig. \ref{fig:potentials}). We term these two different potential types as ``type (I)'' and ``type (II)''  potentials, respectively. 

Setting the righthand side of Eqn. (\ref{FE}) to zero, it is seen that as $d_* \to d_0^-$ and $d_* \to d_0^+$ (for a non-vanishing ratio $a\equiv \frac{d_0}{N}$ in the thermodynamic $N \to \infty$ limit), two transitions appear at, respectively,  $\beta_*^{(1)} = \ln \frac{1-a}{a}$ and $\beta_*^{(\gamma)} =  \frac{\ln\frac{1-a}{a}}{\gamma}$. The existence of these two distinct  transition temperatures is inextricably tied to the two distinct slopes (of values  $1,\gamma$) of the piecewise linear potential of Eqn. (\ref{eq:piecewise}) when $d_* \to d_0^-$ and $d_* \to d_0^+$. In what follows, we will examine, in some detail, what occurs at these two temperatures.

\begin{figure}[!htb]
    \centering
    \includegraphics[width=0.45\textwidth]{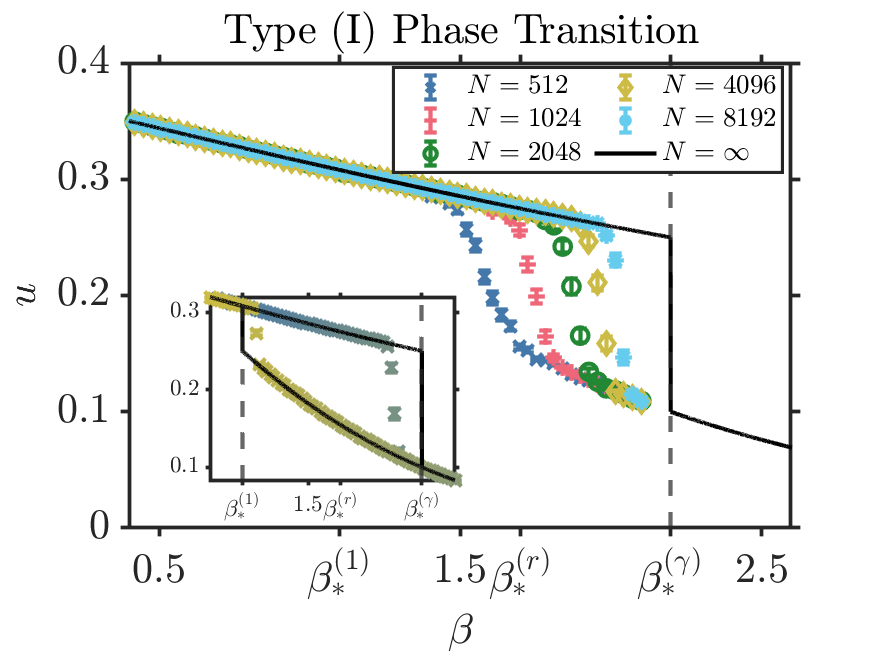}
    \includegraphics[width=0.45\textwidth]{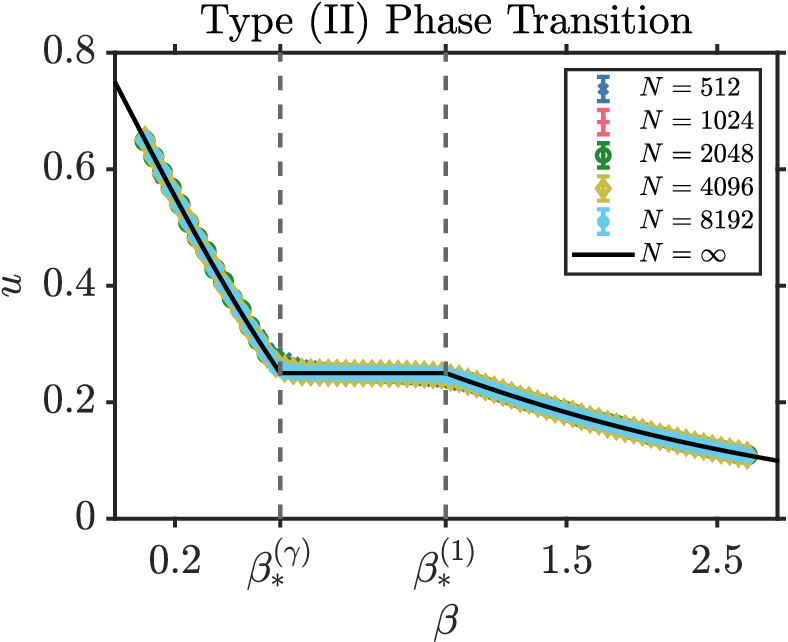}
    \caption{The internal energy density $u$ as a function of the inverse temperature $\beta$ for two different potentials of the form of Eqn. (\ref{eq:piecewise})- one with (a) $\gamma=\frac{1}{2}$ and the other with  (b) $\gamma=2$. Varying sizes $L$ are considered, while $M$ and $a\equiv \frac{d_0}{N}$ are fixed to be $2$ and $\frac{1}{4}$ respectively. Two distinct types of phase transitions are seen. The detailed numerical settings are similar to Fig. \ref{fig:no_phaset}. For (a) $\gamma=\frac{1}{2}$, the inset shows the memory effect of the type (I) potential, simulated with $N=8192$. Scattering points colored blue to yellow depict the entire process from high (random configurations at the very beginning) to low temperature, and then back from low to high temperature. At each temperature, the system underwent $110$ Metropolis-Hasting sweeps from the final configuration from the previous temperature. The whole process can be roughly considered as a counterclockwise `hysteresis'. Note that the position of $\beta_*^{(r)}$ in panel (a) is only roughly marked and is not strictly equal to the actual value.}
    \label{fig:phaset}
\end{figure}

\begin{figure}[!htb]
    \centering
    \includegraphics[width=0.43\textwidth]{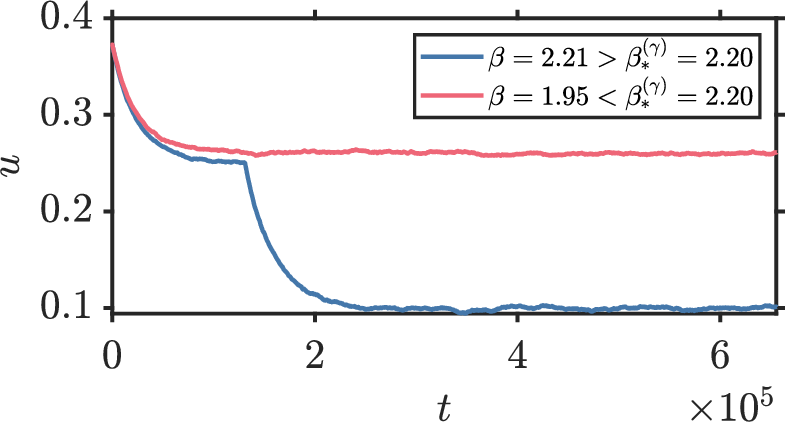}
    \caption{Metropolis-Hasting dynamics on a system with $N=32,768$ and $M=2$. Below (blue) and above (red) the transition temperature, the system converges to different internal energy densities, despite tiny temperature differences. At $\beta=1.95$, the system would be `stuck' at the local free energy minima at $d^{*(\gamma)}=\frac{N}{1+e^{\beta\gamma}}>d_0$ and therefore has difficulty falling into the other local free energy minima at $d^{*(1)}=\frac{N}{1+e^{\beta}}<d_0$. The time $t$ denotes the number of Metropolis steps (number of flipping attempts).}
    \label{fig:dynamics}
\end{figure}

We start by considering ``type (I)'' piecewise linear potential (for which $\gamma<1$) and ask what occurs as the system is heated. At inverse temperatures just above $\beta_*^{\gamma}$, i.e.,  $\beta=\beta_*^{(\gamma)}+\epsilon$ with $0<\epsilon \ll 1$, Eqn. (\ref{FE}) implies that the system is most likely to be at $d_*=\frac{N}{1+\left(\frac{1-a}{a}\right)^{\frac{1}{\gamma}}}-\epsilon'<d_0$ (Eqn. (\ref{eq:freeminima})), where $\epsilon'\equiv \frac{N(\frac{1-a}{a})^{\frac{1}{\gamma}}\epsilon}{(1+(\frac{1-a}{a})^{\frac{1}{\gamma}})^2} \ll N$. However, on further heating, when the inverse temperature is lowered such that  $\beta<\beta_*^{(\gamma)}$, two minima of the free energy of Eqn. (\ref{Fd}) appear: one at $d^{*(\gamma)}=\frac{N}{1+e^{\beta\gamma}}$ and the other at $d^{*(1)}=\frac{N}{1+e^{\beta}}$. The two  minima are  located on opposite sides of $d_0$. Reaching the minimum on the left, namely $d^{*(1)}$ when starting from any random configuration having $d>d_0$ is challenging. Indeed, at an inverse temperature $\beta=(\beta_*^{(\gamma)}-\epsilon)>\beta_*^{(1)}$, the system is most likely to stay at $d_*=(d_0+\epsilon')$, where $\epsilon'=N\gamma a(1-a)\epsilon \ll N$, and therefore will not easily fall into the `inner well'. This state of affairs leads to a discontinuity in the internal energy density $u=\frac{\langle E \rangle}{N}$ at $\beta=\beta_*^{(\gamma)}$, see Fig. \ref{fig:phaset}(a). In Fig. \ref{fig:dynamics}, we depict the Metropolis dynamics\cite{lipowski_heat-bath_2023} around the transition at $\beta_*^{(\gamma)}$. Clearly our ``type (I)'' phase transition is, as befits its name, a discontinuous one.  Note the thermodynamic average here, denoted as $\langle\cdot\rangle$, is not computed in the strict sense of thermodynamic equilibrium, but rather is a statistical average of the systems obtained after performing Monte Carlo simulations starting from random configurations that naturally concentrate on high energies. Conversely, if the systems start from low-energy configurations, as we vary the temperature, the phase transition appears, instead, at $\beta=\beta_*^{(1)}$ (see inset the of  Fig. \ref{fig:phaset} (a)). This dependence on initial conditions is reminiscent of hysteresis and memory effects in spin glasses and other systems\cite{vincent_aging_nodate,noauthor_aging_nodate,baity-jesi_memory_2023}. In fact, there exists a `real' ($r$) inverse transition temperature, $\beta_*^{(1)}<\beta_*^{(r)}<\beta_*^{(\gamma)}$, as the result of the competition between two free energy minima: $F\left(\frac{1}{1+e^{\beta_*^{(r)}\gamma} }\right)=F\left(\frac{1}{1+e^{\beta_*^{(r)}}}\right)$. We next consider using a `real' sampler that draws from the real Boltzmann distribution, as $\beta$ monotonically decreases and crosses $\beta_*^{(r)}$, a jump in internal energy still exists for reasons similar to those discussed for $\beta_*^{(\gamma)}$. For computational convenience associated with the typical launch of the Monte Carlo simulation from random initial configurations, we will employ $\beta_*^{(\gamma)}$ as a proxy for the `real' inverse temperature $\beta_*^{(r)}$ of the actual phase transition\footnote{Determining the precise value of $\beta_{*}^{(r)}$ is also numerically nontrivial. There are inherent difficulties when starting from random configurations and applying Monte Carlo sampling to systems like the Eggbox Ising model that harbor multiple free energy minima.}.

\begin{figure}[!htb]
    \centering
    \includegraphics[width=0.45\textwidth]{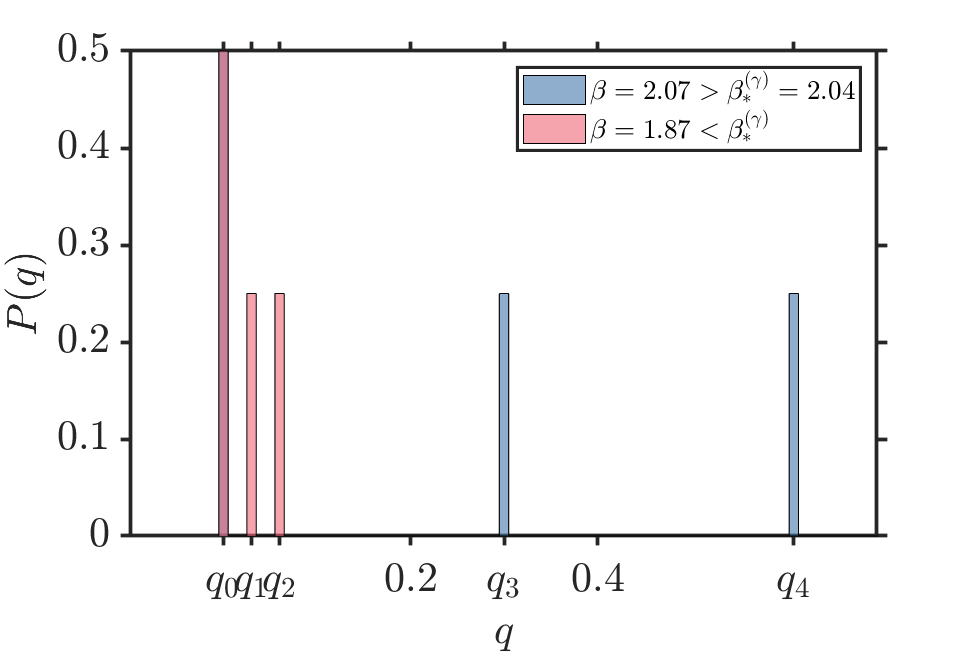}
    \caption{The overlap distribution $P(q)$ for temperatures below and above the transition temperature $\beta_*^{(\gamma)}$, on a single disorder realization with $N=65,536$, $M_0=2$ and $k=2$. We also set $a = \frac{3}{8}$, $\gamma = \frac{1}{4}$. To ensure the replicas reached equilibrium, each underwent $110$ sweeps. The overlap distribution was calculated based on $400$ replicas of a single disorder realization. Specifically, this system possesses $M_02^{k-1}=4$ pure states/local minima. To compute the Parisi overlap $q_{\alpha\beta}=\frac{1}{N}\sum_{i=1}^N\langle s_i\rangle_\alpha\langle s_i \rangle_\beta$, for each pure state, we identified the corresponding replicas (approximately 100 in number) and computed their average to obtain $\langle s_i \rangle_\alpha$. From left to right ($q_0, q_1, q_2$ for $\beta=1.87$ and $q_0, q_3, q_4$ for $\beta=2.07$), the three distribution peaks are associated with the overlaps between (1) replicas in different pure states with different `parents' (Fig. \ref{fig:fractalrsb}); (2) replicas in different pure states with the same parent; (3) the self-replica overlap (amongst states in the same valley).}
    \label{fig:overlapt}
\end{figure}

The ``type (I)'' piecewise linear potential ramp can be employed as a trap to have the system remain in high energy configurations at temperatures just above the transition temperature. As the simplest application of this property, consider a scenario where $a\equiv \frac{d_0}{N}$ is close to $\frac{1}{2}$ and $\beta \to \beta_*^{(\gamma)-}$. In such a case, the system is highly likely to be attracted to high-energy configurations, and sampling of replicas would reveal that their overlaps are almost all $0$. However, when $\beta \to \beta_*^{(\gamma)+}$, the system falls into local minima, leading to a dispersion of the overlap distribution towards non-zero values, see  Fig. \ref{fig:overlapt}. As $\beta$ increases beyond $\beta_*^{(\gamma)}$, the overlaps between the replicas in the same valley and those of replicas in the different valleys belonging to the same `parent' (Fig. \ref{fig:fractalrsb}) rapidly diverge from one another. For instance, it can be readily demonstrated that as $\beta$ asymptotically veers, from the left and right, towards the inverse transition temperature $\beta_*^{(\gamma)}$, the self-overlap $q_{\alpha\alpha}$ respectively approaches $\left(1-\frac{2}{1+e^{\beta_*^{(\gamma)}\gamma}}\right)^2$ and $\left(1-\frac{2}{1+e^{\beta_*^{(\gamma)}}}\right)^2$.

\begin{figure}
    \centering
    \begin{subfigure}{0.40\textwidth}
    \includegraphics[width=\textwidth]{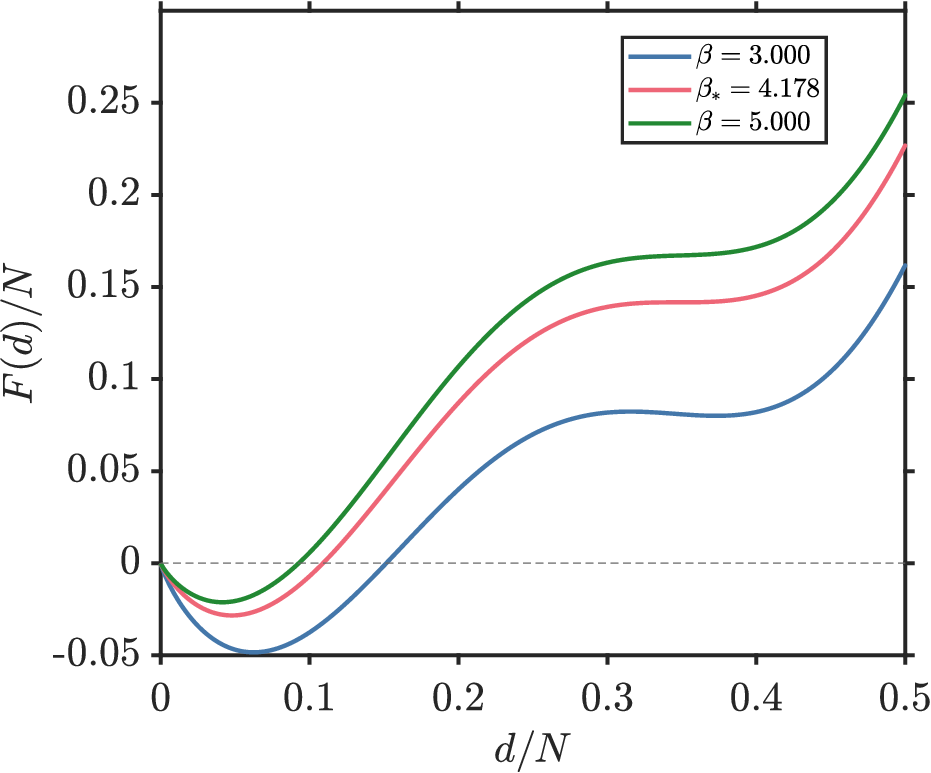}
    \caption{}
    \end{subfigure}
    \begin{subfigure}{0.40\textwidth}
    \includegraphics[width=\textwidth]{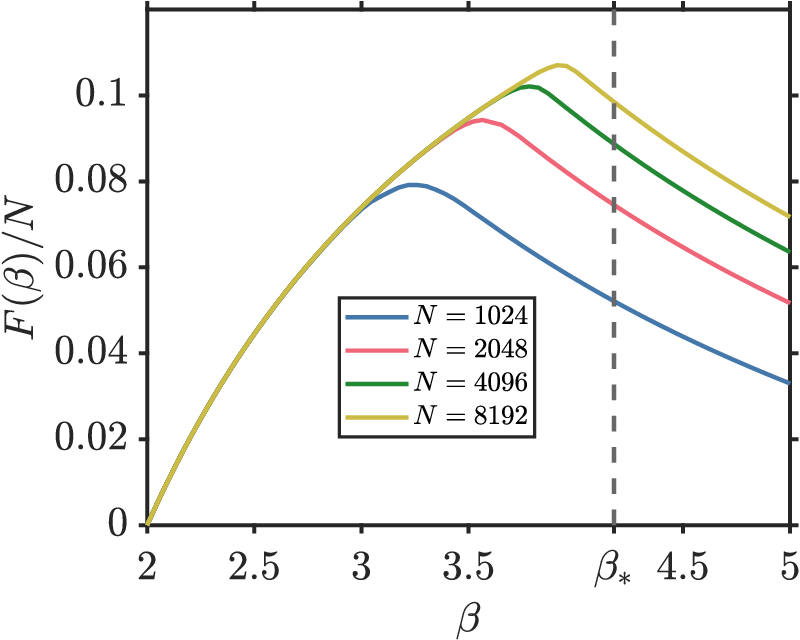}
    \caption{}
    \end{subfigure}
    \caption{(a) The free energy $F(d)$ of the potential $V(d)= \frac{\pi d^2}{2N}-\frac{N}{4\pi}\cos\left( \frac{4\pi d}{N}\right)$ and its derivative $F'(d)$ (inset). For simplicity, we set $F(0)=0$ and normalized by $N$. We evaluate $F(d)$ by integrating the righthand side of Eqn. (\ref{FE}). When the inverse temperature $\beta$ drops below $\beta_*=4.178$ (higher yellow curve to lower blue curve), a new free energy minimum appears at $\frac{d}{N} \simeq 0.35$. This leads to a phase transition at $\beta_*$. (b) The normalized free energy $F(\beta)/N$ computed by Monte Carlo simulation (similar in Fig. \ref{fig:no_phaset}) via an integration of $u(\beta)=\frac{d(\beta F(\beta))}{d\beta}$. Consistent with the displayed free energy density traces for increasing system size, $F'(\beta)/N$ may be discontinuous at a phase transition at $\beta=\beta_*$ as $N\to\infty$.}
    \label{fig:sin_freeenergy}
\end{figure}

\begin{figure}[!htb]
    \centering
    \includegraphics[width=0.42\textwidth]{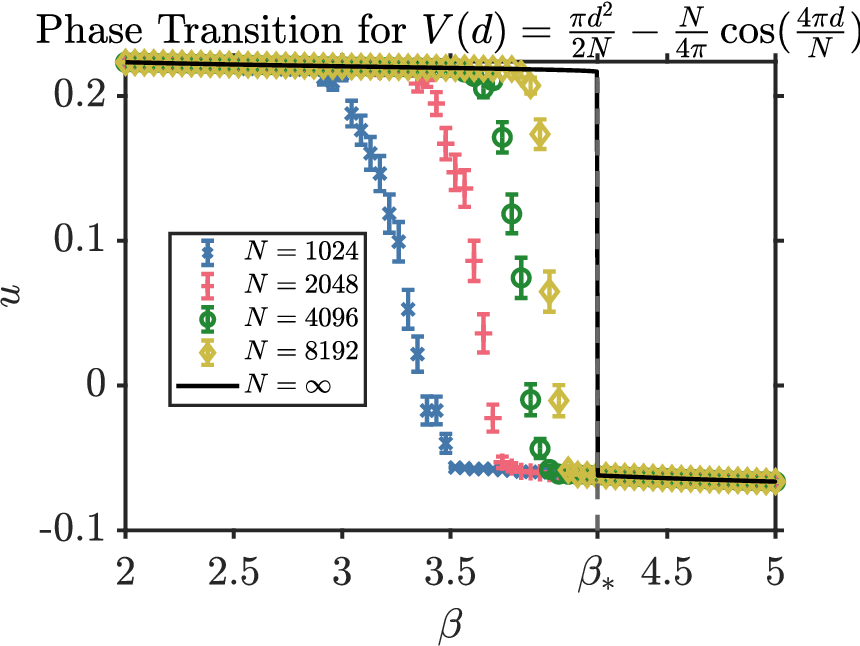}
    \caption{The internal energy density $u$ as a function of the inverse temperature computed for the  $V(d)= \frac{\pi d^2}{2N}-\frac{N}{4\pi}\cos( \frac{4\pi d}{N})$. Similar to Fig. \ref{fig:phaset} for the rather different piecewise linear $V$ of Eqn. (\ref{eq:piecewise}), in the thermodynamic limit, a first-order phase transition also  appears here (at $\beta_*=4.178$).}
    \label{fig:phaset_cont}
\end{figure}

We reiterate that although we adopted a piecewise linear potential for analytical transparency and a sharp physical picture, it is not merely a ``toy'' example. Rather, it distills the generic mechanism behind the discontinuous transitions discussed above: the competition between multiple local minima of the free energy $F(d)$ (equivalently, multiple stable solutions of the stationarity condition) and the resulting metastability and hysteresis under realistic dynamical protocols. This mechanism persists for a broad class of more ``natural'' and smooth choices of $V(d)$, and one may explicitly construct continuous potentials that reproduce the same two-minima structure. For instance, for $V(d)= \frac{\pi d^2}{2N}-\frac{N}{4\pi}\cos\left( \frac{4\pi d}{N}\right)$ (see the rightmost panel of Fig. \ref{fig:potentials}), a type (I) transition appears at $\beta_*=4.178$: as $\beta$ drops below $\beta_*$, a new free-energy minimum emerges at $\frac{d}{N} \simeq 0.35$ (see Fig. \ref{fig:sin_freeenergy}), giving rise to a first-order transition (Fig. \ref{fig:phaset_cont}).

\begin{figure}[!htb]
    \centering
    \includegraphics[width=0.45\textwidth]{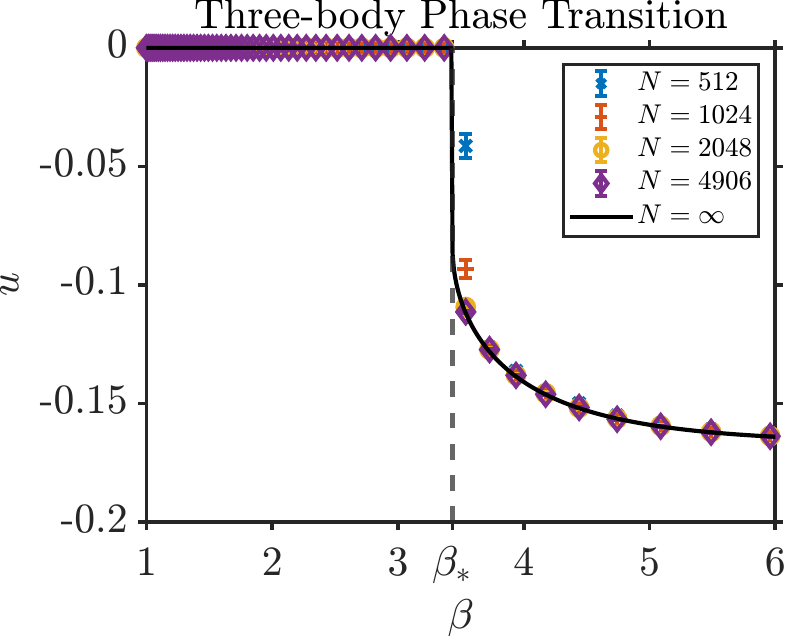}
    \caption{Phase transition in the fully-connected ferromagnetic three-body Ising model under a slow heating protocol. The mechanism mirrors the type (I) Eggbox transition: two competing free-energy minima coexist over a temperature window, producing metastability and a discontinuous jump when a metastable minimum vanishes.}
    \label{fig:threebody_hysteresis}
\end{figure}

\paragraph{Three-body ferromagnetic Ising model.}
The same ``multiple-minima competition'' mechanism also appears in a more microscopic and widely studied setting, namely the fully-connected ferromagnetic three-body Ising model. Consider
\begin{equation}
\label{eq:threebody_H}
E(\boldsymbol{\sigma})=
-\frac{J}{N^{2}}\sum_{1\le i<j<k\le N}\sigma_i\sigma_j\sigma_k,
\qquad \sigma_i\in\{\pm 1\},\ J>0.
\end{equation}
Let $d$ denote the Hamming distance to a fully polarized configuration (equivalently, the number of spins flipped relative to it, so that $\sum_i\sigma_i=N-2d$). A straightforward combinatorial evaluation yields an energy that depends only on $d$,
\begin{equation}
\label{eq:threebody_Ed_exact}
E(d)= -\frac{J}{6N^{2}}\Big[(N-2d)^{3}-(3N-2)(N-2d)\Big].
\end{equation}
For $d$ not too large compared to $N$, the subleading linear term may be neglected and
\begin{equation}
\label{eq:threebody_Ed_approx}
E(d)\simeq -\frac{J}{6N^{2}}(N-2d)^{3}.
\end{equation}
Substituting Eq.~(\ref{eq:threebody_Ed_approx}) into the discrete stationarity condition Eq.~(\ref{FE}) (and using $d+1\simeq d$ for large $N$) gives
\begin{equation}
\label{eq:threebody_FE}
\frac{J(N-2d)^{2}}{N^{2}}-\frac{1}{\beta}\ln\!\left(\frac{N-d}{d}\right)\simeq 0.
\end{equation}
In terms of $\alpha\equiv d/N$, this becomes the transcendental equation
\begin{equation}
\label{eq:threebody_transcend}
\beta J(1-2\alpha)^{2}-\ln\!\left(\frac{1-\alpha}{\alpha}\right)=0.
\end{equation}
Numerically, for $J=1$ and sufficiently large $\beta$, Eq.~(\ref{eq:threebody_transcend}) admits two solutions: one ferromagnetic branch with $\alpha \frac{1}{2}$ and one disordered branch with $\alpha= \frac{1}{2}$. Upon heating under local-update dynamics, the system may remain trapped on the ferromagnetic branch until a inverse transition temperature $\beta=\beta_{*}=3.433$ at which the ferromagnetic local minimum disappears, leading to a rapid jump to $\alpha=\frac{1}{2}$, see Fig.~\ref{fig:threebody_hysteresis}.

We now turn to the ``type (II)'' piecewise linear potentials with $\gamma>1$. Similar to our above discussion for the ``type (I)'' potential, we scan all non-negative temperatures and vary $\beta$ from $\infty$ to $0$. When $\beta=\beta_*^{(1)}$, the system `climbs' to $d_*=d_0$. However, since the slope $\gamma>1$, the system can only continue to `climb' when $\beta < \beta_*^{(\gamma)}$. Therefore, the system possesses two discontinuities in the heat capacity, and both $\beta_*^{(1)}$ and $\beta_*^{(\gamma)}$ appear as inverse transition temperatures (see Fig. \ref{fig:phaset}(b)).

To summarize, by manipulating the potential function $V(d)$ and changing parameters such as $M$ and $k$, one can create and study models exhibiting different desired types of phase transitions.

\section{Additional Variants}

Our model is highly tunable and readily accessible to numerical investigation. The complex energy landscape is associated with various  physical behaviors. Further complexity can be introduced along various interrelated lines. First, the local minima $\{\epsilon_\alpha\}$ may be made different and adjusted so as to create a global minimum along with other local minima can be considered. Second, the potential $V(d)$ itself can be endowed with multiple local minima by, e.g., choosing it to be a function such as $V(d) = d + \cos(d)$. Third, establishing bridges between various local minima $\{\boldsymbol{\xi}^\alpha\}$ would facilitate the system configuration to traverse more conveniently among these valleys. For example, we can write down an effective free energy for the Eggbox Ising model,
\begin{equation}\label{eq:model_energy_sa}
F^{\rm eff}_\beta(\boldsymbol{\sigma}) =\epsilon_{\alpha_0} + V( d(\boldsymbol{\sigma},\boldsymbol{\xi}^{\alpha_0}))+\lambda_\beta \sum_\alpha d(\boldsymbol{\sigma},\boldsymbol{\xi}^\alpha).
\end{equation}
Here, $\alpha_0\equiv \arg\min\limits_\alpha d(\boldsymbol{\sigma},\boldsymbol{\xi}^\alpha)$ marks the index of the local minima that is closest to $\boldsymbol{\sigma}$. The addition of the term $\sum_\alpha d(\boldsymbol{\sigma},\boldsymbol{\xi}^\alpha)$- the sum of the distances from $\boldsymbol{\sigma}$ to all local minima $\{\boldsymbol{\xi}^\alpha\}$- favors that the system approaches each configuration  $\boldsymbol{\xi}^\alpha$ as closely as possible. In other words, the configuration of the system is attracted to a subspace `spanned' by the local minima thereby constraining it from indiscriminately traversing the entire space. The prefactor $\lambda_\beta$ is temperature dependent and here we set it to be $\lambda_\beta = e^{-\beta}$. At zero temperature, the prefactor vanishes, and the system's energy local minima revert to $\{\boldsymbol{\xi}^\alpha\}$.

\begin{figure}[!htb]
    \centering
    \includegraphics[width=0.45\textwidth]{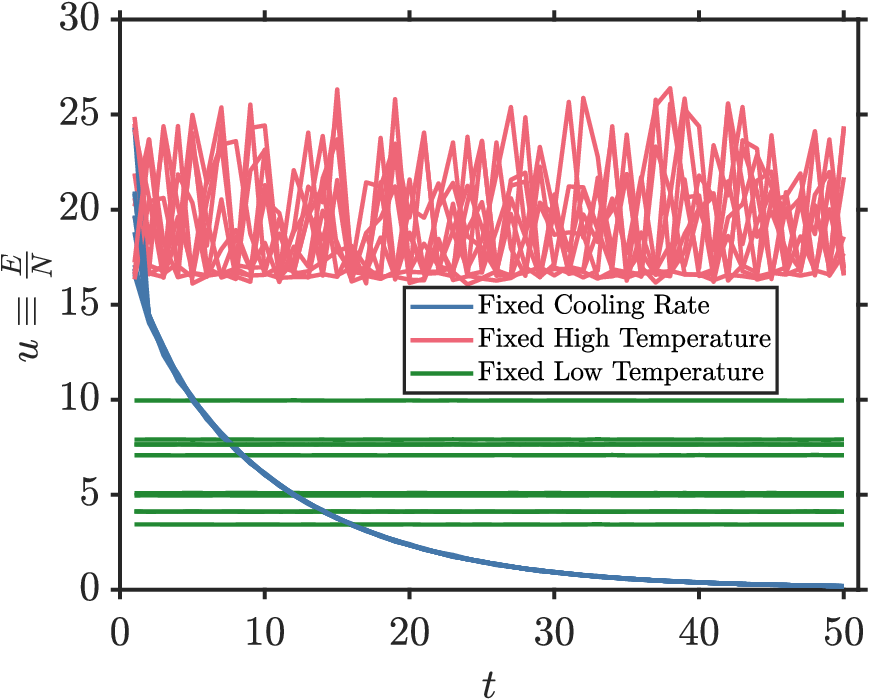}
    \includegraphics[width=0.45\textwidth]{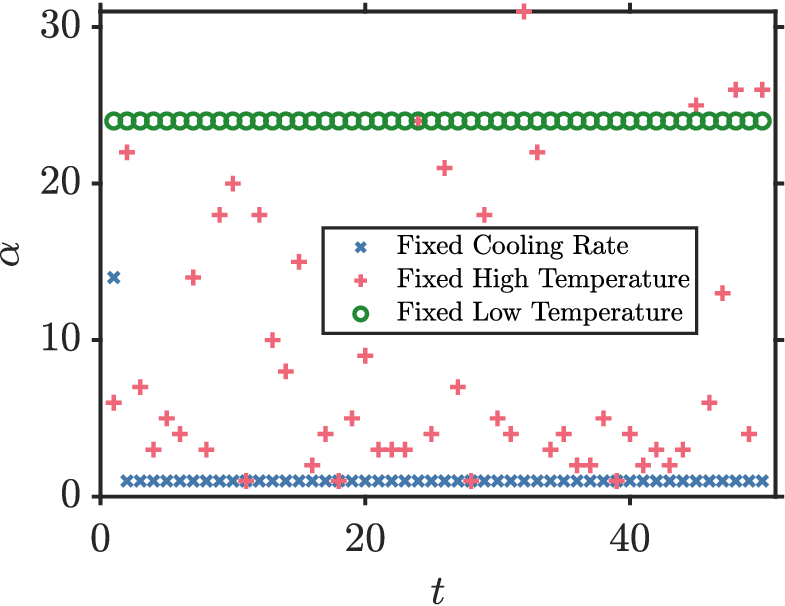}
    \caption{(a) For an $N=256,M=32$ system described by Eqn. (\ref{eq:model_energy_sa}), the simulated annealing algorithm was applied for $10$ independent trials for a single disorder realization. Three inverse temperature time  sequences $\{\beta_t\},t=1,2, \dots, 50$ were examined. The first sequence termed the fixed cooling rate sequence, consists of $50$ inverse temperatures equally spaced between $0.001$ and $5.000$. The second and third sequences correspond to $50$ fixed inverse temperatures of $0.001$ or $5.000$ respectively, and are referred to as the fixed high temperature sequence and the fixed low temperature sequence. The fixed cooling rate sequence uniquely precisely located the correct valley $\alpha=1$ at the high-temperature phase and maintained a continuous energy descent during the low-temperature phase. Consequently, the final energy always stood as the lowest among the three sequences. Conversely, the fixed high temperature sequence incessantly traversed among various valleys, while the fixed low temperature sequence predominantly engaged in energy descent within the valley in which it originally commenced. (b) For a single trial, we depicted the variation of the valley index $\alpha$ in which the system configuration resides.}
    \label{fig:satest}
\end{figure}

The Eggbox Ising model possesses an exceedingly rare characteristic, namely the ability to ascertain to which valley of a local minima $\boldsymbol{\xi}^\alpha$ any given system configuration belongs. Taking Simulated Annealing, as referenced in \cite{wang_comparing_2015}, as an illustrative example. If simulated annealing is applied on the Eggbox Ising model, it becomes feasible to observe which temperature sequence\cite{mills_finding_2020} or sweep strategy\cite{isakov2015optimised} aids the system configuration in efficiently exiting the valleys of local minima and transitioning towards the valley of the global minimum. A straightforward numerical simulation was conducted to validate the proposed concept. In this simulation, we sampled $\epsilon_\alpha$ in Eqn. (\ref{eq:model_energy_sa}) uniformly from the interval $[0,10]$, multiplied by the system size $N$. We subsequently rearranged $\{\epsilon_\alpha \}$ to ensure $\epsilon_1<\epsilon_2<\dots<\epsilon_M$, therefore the true system ground state locates at $\boldsymbol{\xi}^1$. We also set $V(d)=d$, again, for simplicity. The outcomes revealed that, for such a system configuration, a temperature sequence transitioning from high to low at a fixed rate typically outperforms both a fixed low temperature sequence and a fixed high temperature sequence, see Fig. \ref{fig:satest} and its caption for detailed illustrations. For most of the cases, only the fixed cooling rate sequence initially precisely identified the correct valley $\alpha=1$ and continuously decreased the energy during the low-temperature phase. Therefore, the final energy was always the lowest among the three sequences. In contrast, the fixed high temperature sequence continually oscillated between different valleys, while the fixed low temperature sequence predominantly underwent energy descent within the initial valley it started in. 

\section{Conclusion and Discussion}

In the current work, we introduced the Eggbox Ising model and elucidated some of its basic properties. We discussed, how various parameters affect its physical characteristics, such as the RSB structures (Fig. \ref{fig:fractalrsb}) and the density of states (Fig. \ref{fig:dos}) determined by $M_0,c,k$; or phase transitions (Fig. \ref{fig:no_phaset},\ref{fig:phaset},\ref{fig:dynamics},\ref{fig:phaset_cont},\ref{fig:overlapt}) induced by different potentials $V(d)$ (Fig. \ref{fig:potentials}). Our framework enables the systematic construction of models that have any given overlap distribution $P(q)$. This demonstrates the considerable potential of the Eggbox Ising model as a framework for investigating numerous physical phenomena, particularly in disordered systems. On the other hand, it can enhance our understanding of numerical algorithms such as Simulated Annealing (Fig. \ref{fig:satest}), thereby aiding in the optimization and even the design of these algorithms.

Following this, we aim to specifically discuss one additional aspect: the connections between the Eggbox Ising model and other models.

\subsection{Relation to Other Models}
\label{sec:relation}
We now briefly discuss several models that share some similarities with ours.

$\bullet$ The {\it (Field) Eggbox Model} \cite{massen_exploring_2007}. This model features an energy landscape that is similar to that of our Eggbox Ising model. The distinction lies in this model's energy as defined on a Euclidean space, thereby possessing a clear geometric structure. Due to the constraints imposed by the us of Euclidean space, this model did not feature a fractal $k$-RSB structure with ultrametricity.

$\bullet$ The {\it Random Energy Model} (REM)\cite{gross_simplest_1984} and the {\it Generalized Random Energy Model} (GREM) \cite{jana_generalized_2006}. Compared to the REM, the energies between different spin configurations in the GREM are correlated, and the strength of this correlation is proportional to their similarity (for example, the overlap). In this regard, the GREM shares a conceptual resemblance with the Eggbox Ising model, yet they diverge fundamentally. First, in the Eggbox Ising model, the configuration energies in $\mathcal{LM}$ are highly correlated, despite potentially significant Hamming distances. Secondly, the Eggbox Ising model delineates a clearer energy landscape compared to the GREM, with the energies of adjacent spin configurations governed by a clearly defined function as established in this paper. It should be noted, however, that introducing randomness into $V(d)$ may, to an extent, emulate the behavioral characteristics of the GREM.

$\bullet$ The {\it Random Code Ensemble} (RCE) \cite{mezard2009information}.
As we mentioned in the current work, the notion of $\mathcal{LM}$ within the Eggbox Ising model bears a conceptual similarity with the Codebook in the RCE model. Despite this similarity, there exist essential distinctions between the two models. In the context of the RCE model, the primary focus is on the recovery of the Codebook element $\boldsymbol{\sigma}$ upon receiving the channel output $\boldsymbol{y}$. Consequently, the system's configuration space is the Codebook itself, a characteristic that markedly differentiates it from the Eggbox Ising model.

$\bullet$ The {\it Hopfield Model} \cite{hopfield1982neural}. In the absence of an external field, the Hopfield model energy $E=-\sum_{ij}\omega_{ij}s_is_j$ where $\omega_{ij}=\sum_{\xi=1}^{M}V^\xi_iV^\xi_j$ with $V^\xi$ denotes a specific pattern to be stored. The energy minima of the Hopfield model usually restores the pattern $V^\xi$. However, compared to the Eggbox Ising model, the overall energy landscape is not transparent.

{\bf Acknowledgements.} ZN is grateful to the Leverhulme-Peierls senior researcher Professorship at Oxford supported by a Leverhulme Trust International Professorship grant [number LIP-2020-014]. YX acknowledges support from the NSF through OAC-2118310.

\bibliography{ref}
\bibliographystyle{unsrt}

\end{document}